# Extended Solar Emission – an Analysis of the EGRET Data

Elena Orlando, Dirk Petry and Andrew Strong

*Max-Planck-Institut für extraterrestrische Physik, Postfach 1312, D-85741 Garching, Germany*

**Abstract.** The Sun was recently predicted to be an extended source of gamma-ray emission, produced by inverse-Compton scattering of cosmic-ray electrons with the solar radiation. The emission was predicted to contribute to the diffuse extragalactic background even at large angular distances from the Sun. While this emission is expected to be readily detectable in future by GLAST, the situation for available EGRET data is more challenging. We present a detailed study of the EGRET database, using a time dependent analysis, accounting for the effect of the emission from 3C 279, the moon, and other sources, which interfere with the solar signal. The technique has been tested on the moon signal, with results consistent with previous work. We find clear evidence for emission from the Sun and its vicinity. The observations are compared with our model for the extended emission.

**Keywords:** Cosmic rays, gamma-ray emission, Sun, EGRET.
**PACS:** 95.85, 96.50, 96.60

## THE EXTENDED SOLAR EMISSION MODEL

The heliosphere has been studied as an extended source of gamma-ray emission, produced by inverse-Compton scattering of cosmic-ray electrons with the solar photon field [1,2]. For this analysis our model [1] has been improved using the modulated electron spectrum at all distances following [2] instead of the measured local electron spectrum, and using the anisotropic inverse-Compton scattering formalism [3].

## ANALYSIS OF THE EGRET DATA

We analyzed the EGRET data using the code developed for the moving target Earth [4] and adding necessary features (solar and lunar ephemerides, occultation, background point source trace calculations). The diffuse background was reduced by excluding the Galactic plane. Otherwise all available exposure within mission phase 1-3 was used. When the Sun passed by other gamma-ray sources (moon, 3C 279 and several quasars), these sources were included in the analysis. Details will be given in [5].

We fitted the data in the Sun-centred system using a multi-parameter likelihood fitting technique, leaving as free parameters the solar extended inverse-Compton flux from the model, the solar disk flux from pion decay [7], a uniform background, and the flux of 3C279, the dominant background point source. The moon flux was determined from moon-centred fits and the 3EG source fluxes were fixed at their catalogue values. All components were convolved with the energy-dependent EGRET PSF. The region used for fitting is a circle of radius 10º centred on the Sun. Since the interesting parameters are solar disk source and extended emission, the likelihood is maximized over the other components. In order to verify our method, we checked that the fluxes of the Crab Nebula, 3C 279, and in particular the moon [6] were reproduced.

### Results

The log-likelihood ratio for E >100 MeV is displayed in Fig.1 as a function of solar disk flux and extended flux, compared with the model prediction of solar inverse-Compton flux for modulation parameter 500 MV at 1 AU. The solar emission is detected with 5.3σ significance. There is evidence for the extension of the emission at a level of 2.7σ; the maximum log L indicates an extended component with a flux compatible with the IC model. The total flux from the Sun is more than expected for the disk source [7], so this is clear evidence for the IC emission even without the proof of extension. We find that the measured extended flux is fully consistent with the model. Figure 2 shows the fitted model counts of the main components and the total including uniform background.

In future work we will perform a detailed spectral analysis, refine the analysis of systematic errors, and study different models for the modulated electron spectrum. This is important for future missions such as GLAST and for studying solar modulation.

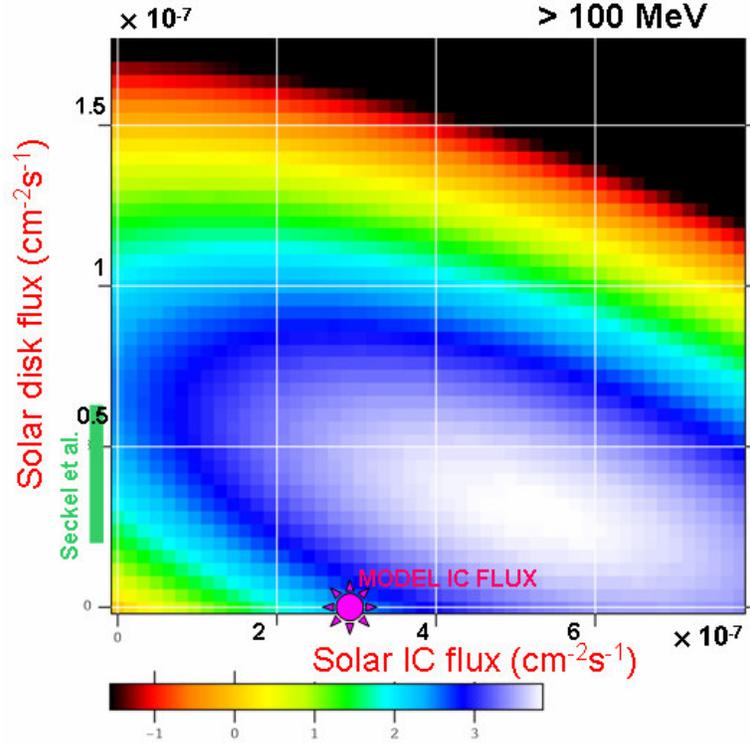

**FIGURE 1.** Log Likelihood above 100 MeV as function of the solar disk flux and extended solar flux, relative to point at (0,0). The level of our predicted IC model flux and the predicted disk flux [7] are shown.

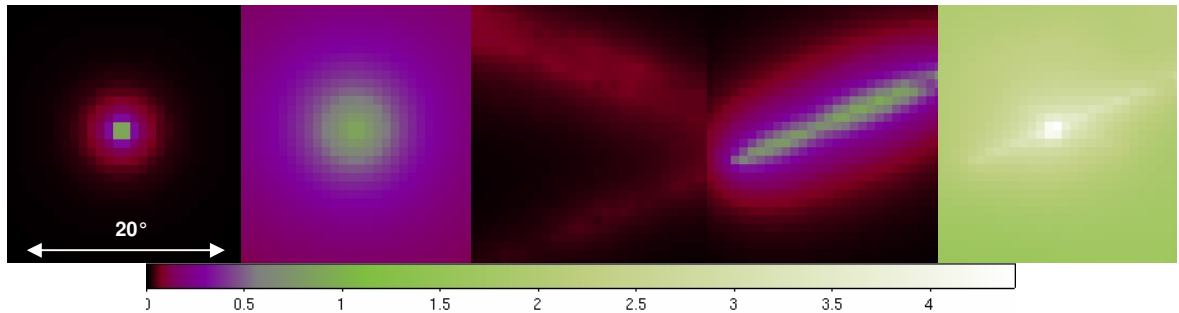

**FIGURE 2.** Fitted model counts of the main components centered on the Sun. From left to right: Sun disk, Sun IC, moon, 3C 279, and the total predicted counts including uniform background. The colors show the counts/pixel, for $0.5° \times 0.5°$ pixels.